\input jytex.tex   
\typesize=10pt \magnification=1200 \baselineskip17truept
\footnotenumstyle{arabic} \hsize=6truein\vsize=8.5truein

\sectionnumstyle{blank}
\chapternumstyle{blank}
\chapternum=1
\sectionnum=1
\pagenum=0

\def\begintitle{\pagenumstyle{blank}\parindent=0pt
\begin{narrow}[0.4in]}
\def\endtitle{\end{narrow}\newpage\pagenumstyle{arabic}}


\def\beginexercise{\vskip 20truept\parindent=0pt\begin{narrow}[10
truept]}
\def\endexercise{\vskip 10truept\end{narrow}}


\def\eql#1{\eqno\eqnlabel{#1}}
\def\ref{\reference}
\def\peq{\puteqn}
\def\pref{\putref}

\def\mgn{\marginnote}
\def\bex{\begin{exercise}}
\def\eex{\end{exercise}}


\font\open=msbm10 


\def\StretchRtArr#1{{\count255=0\loop\relbar\joinrel\advance\count255 by1
\ifnum\count255<#1\repeat\rightarrow}}
\def\StretchLtArr#1{\,{\leftarrow\!\!\count255=0\loop\relbar
\joinrel\advance\count255 by1\ifnum\count255<#1\repeat}}

\def\StretchLRtArr#1{\,{\leftarrow\!\!\count255=0\loop\relbar\joinrel\advance
\count255 by1\ifnum\count255<#1\repeat\rightarrow\,\,}}

\def\mbox#1{{\leavevmode\hbox{#1}}}

\def\hspace#1{{\phantom{\mbox#1}}}
\def\oZ{\mbox{\open\char90}}

\def\al{\alpha}
\def\be{\beta}
\def\ga{\gamma}

\def\Ga{\Gamma}

\def\ep{\epsilon}

\def\la{\lambda}

\def\th{\theta}

\def\ze{\zeta}

\def\caD{{\cal D}}

\def\zf{$\zeta$--function}
\def\zfs{$\zeta$--functions}


\def\frac#1/#2{\leavevmode\kern.1em
\raise.5ex\hbox{\the\scriptfont0 #1}\kern-.1em/\kern-.15em
\lower.25ex\hbox{\the\scriptfont0 #2}}
\def\sfrac#1/#2{\leavevmode\kern.1em
\raise.5ex\hbox{\the\scriptscriptfont0 #1}\kern-.1em/\kern-.15em
\lower.25ex\hbox{\the\scriptscriptfont0 #2}}

\def\gtorder{\mathrel{\raise.3ex\hbox{$>$}\mkern-14mu
             \lower0.6ex\hbox{$\sim$}}}
\def\ltorder{\mathrel{\raise.3ex\hbox{$<$}\mkern-14mu
             \lower0.6ex\hbox{$\sim$}}}

\def\semidirprod{\rlap{\ss C}\raise1pt\hbox{$\mkern.75mu\times$}}
\def\for{\lower6pt\hbox{$\Big|$}}
\def\fish{\kern-.25em{\phantom{abcde}\over \phantom{abcde}}\kern-.25em}


\def\boxit#1{\vbox{\hrule\hbox{\vrule\kern3pt
        \vbox{\kern3pt#1\kern3pt}\kern3pt\vrule}\hrule}}
\def\dalemb#1#2{{\vbox{\hrule height .#2pt
        \hbox{\vrule width.#2pt height#1pt \kern#1pt \vrule
                width.#2pt} \hrule height.#2pt}}}

\def\ol{\overline}
\def\frac#1#2{{{#1}\over{#2}}}


\def\cosec{{\rm cosec\,}}
\def\etc{{\it etc.}}

\def\eg{{\it e.g.}}
\def\ie{{\it i.e. }}
\def\cf{{\it cf }}
\def\pa{\partial}


\def\sumdasht#1#2{{\mathop{{\sum}'}_{#1}^{#2}}}

\def\3j#1#2#3#4#5#6{\left\lgroup\matrix{#1&#2&#3\cr#4&#5&#6\cr}
\right\rgroup}

\def\man{{\cal M}}

\def\m?{\mgn{?}}

\def\pa{\partial}

\def\beq{\begin{eqnarray}}
\def\eeq{\end{eqnarray}}


\def\aop#1#2#3{{\it Ann. Phys.} {\bf {#1}} ({#2}) #3}

\def\cmp#1#2#3{{\it Comm. Math. Phys.} {\bf {#1}} ({#2}) #3}
\def\cqg#1#2#3{{\it Class. Quant. Grav.} {\bf {#1}} ({#2}) #3}

\def\jmp#1#2#3{{\it J. Math. Phys.} {\bf {#1}} ({#2}) #3}
\def\jpa#1#2#3{{\it J. Phys.} {\bf A{#1}} ({#2}) #3}

\def\np#1#2#3{{\it Nucl. Phys.} {\bf B{#1}} ({#2}) #3}
\def\pl#1#2#3{{\it Phys. Lett.} {\bf {#1}} ({#2}) #3}

\def\prp#1#2#3{{\it Phys. Rep.} {\bf {#1}} ({#2}) #3}
\def\pr#1#2#3{{\it Phys. Rev.} {\bf {#1}} ({#2}) #3}

\def\prD#1#2#3{{\it Phys. Rev.} {\bf D{#1}} ({#2}) #3}

\def\rmp#1#2#3{{\it Rev. Mod. Phys.} {\bf {#1}} ({#2}) #3}

\def\zfp#1#2#3{{\it Z. f. Phys.} {\bf {#1}} ({#2}) #3}

\def\cras#1#2#3{{\it Comptes Rend. Acad. Sci. (Paris)} {\bf{#1}} (#2) #3}
\def\prs#1#2#3{{\it Proc. Roy. Soc.} {\bf A{#1}} ({#2}) #3}

\def\amsh#1#2#3{{\it Abh. Math. Sem. Ham.} {\bf {#1}} ({#2}) #3}
\def\am#1#2#3{{\it Acta Mathematica} {\bf {#1}} ({#2}) #3}
\def\aim#1#2#3{{\it Adv. in Math.} {\bf {#1}} ({#2}) #3}
\def\ajm#1#2#3{{\it Am. J. Math.} {\bf {#1}} ({#2}) #3}

\def\aom#1#2#3{{\it Ann. of Math.} {\bf {#1}} ({#2}) #3}
\def\cjm#1#2#3{{\it Can. J. Math.} {\bf {#1}} ({#2}) #3}

\def\dmj#1#2#3{{\it Duke Math. J.} {\bf {#1}} ({#2}) #3}
\def\invm#1#2#3{{\it Invent. Math.} {\bf {#1}} ({#2}) #3}

\def\jdg#1#2#3{{\it J. Diff. Geom.} {\bf {#1}} ({#2}) #3}

\def\jram#1#2#3{{\it J. f. reine u. Angew. Math.} {\bf {#1}} ({#2}) #3}
\def\jims#1#2#3{{\it J. Indian. Math. Soc.} {\bf {#1}} ({#2}) #3}
\def\jlms#1#2#3{{\it J. Lond. Math. Soc.} {\bf {#1}} ({#2}) #3}

\def\ma#1#2#3{{\it Math. Ann.} {\bf {#1}} ({#2}) #3}

\def\mz#1#2#3{{\it Math. Zeit.} {\bf {#1}} ({#2}) #3}
\def\ojm#1#2#3{{\it Osaka J.Math.} {\bf {#1}} ({#2}) #3}

\def\plms#1#2#3{{\it Proc. Lond. Math. Soc.} {\bf {#1}} ({#2}) #3}
\def\pgma#1#2#3{{\it Proc. Glasgow Math. Ass.} {\bf {#1}} ({#2}) #3}
\def\qjm#1#2#3{{\it Quart. J. Math.} {\bf {#1}} ({#2}) #3}

\def\rmjm#1#2#3{{\it Rocky Mountain J. Math.} {\bf {#1}} ({#2}) #3}

\def\tams#1#2#3{{\it Trans.Am.Math.Soc.} {\bf {#1}} ({#2}) #3}

\begin{title}
\vglue 1truein
\vskip15truept
\centertext {\Bigfonts \bf Determinants on lens spaces}
\vskip 5truept
\centertext {\Bigfonts \bf and cyclotomic units}
\vskip10truept
\centertext{\Bigfonts \bf } \vskip 20truept
\centertext{J.S.Dowker\footnote{dowker@a35.ph.man.ac.uk}} \vskip 7truept
\centertext{\it Department of Theoretical Physics,} \centertext{\it The
University of
 Manchester,} \centertext{\it Manchester, England}
\vskip40truept
\begin{narrow}
The Laplacian functional determinants for conformal scalars and coexact
one--forms are evaluated in closed form on inhomogeneous lens spaces of
certain orders, including all odd primes when the essential part of the
expression is given, formally, as a cyclotomic unit.
\end{narrow}
\vskip 5truept
\vskip 60truept
\vfil
\end{title}
\pagenum=0
\newpage

\section{\bf 1. Introduction.}
Given the solution of a spectral problem, for some differential operator
say, the calculation of the corresponding functional determinant could be
regarded as just a computational challenge but there are, of course, uses
for such objects. In physics they determine the one--loop effective action.
In mathematics, for the de Rham complex, they occur in the analytic torsion,
and elsewhere.

It is not necessary to have the spectrum explicitly available in order to
calculate the determinant, but it helps. For this reason many discussions
revolve around exactly solvable cases and prominent amongst these are the
spheres. Some relevant brief history was attempted in [\pref{DandK5}] so
nothing more will be said on this, just now.

The intention of this paper is to present a small contribution to the
general store of knowledge about spherical determinants, in particular on
lens spaces. These have played an important part in discussions of analytic
torsion, [\pref{Ray}].

In an earlier work, [\pref{dowsut}], amongst other things, I calculated the
determinants on the even homogeneous lens spaces, S$^3/\oZ_q$, (see also
[\pref{NandO}]). In the present work I turn to the inhomogeneous case.
Curiously the results turn out to look quite different when $q$ is an odd
prime. A variety of approaches is offered and I also make some further
technical analysis of the homogeneous case.

\section{2. \bf Lens space spectrum.}

Thinking of the lens space as the spatial section of an Einstein Universe,
I consider only the conformal scalar and divergenceless Maxwell vector
(coexact $1$--form) eigenproblems in computational detail. There is a lot of
prior mathematical work on the spectral problem, as it is relevant for the
analytic torsion (the minimal scalar is needed here) and the $\eta$
invariant, but I shall take the spectral
details as given in the earlier works, [\pref{DandB,DandJ,dowded}], since
they are in the form which I wish to use.

The differential operators under consideration are the scalar Laplacian,
with an addition to make it conformally invariant (in four dimensions), the
de Rham Laplacian, and, for spin--half, the square of the Dirac operator.

The fact that the spectrum is composed of squares of integers (up to a
scaling) means that number theory is almost bound to appear somewhere in the
story, and this will happen.

The eigenfunctions are labelled by two angular momenta, $L$ and $J$ where
${\bf J=L+j}$ and ${\bf j}$ is the spin of the field ($j=0,1/2,1$). The
massless polarization conditions produce the restriction $J=L\pm j$, making
$L$ ultimately a sufficient label.

The spectrum on the full sphere of radius $a$ is,
  $$\eqalign{
  \la^+_{n}&={1\over a^2}\,(j+n)^2,\quad n=1,2,\ldots\,,\quad (J=L+j)\cr
  \la^-_{n}&={1\over a^2}\,(j-n)^2,\quad n=2j+1,2j+2,\ldots\,,\quad(J=L-j)\,.
  }
  \eql{spec}
  $$
where $n=2L+1\in\oZ$ and the spectrum has been split into the parts that
arise from the positive and the negative spectra of the corresponding first
order (pseudo)--operators. If these parts are to be united, it is necessary
to distinguish $j=0$ from $j>0$. The positive and negative eigenfunctions
are related by the interchange, $L\leftrightarrow J$, a parity
transformation. The point is that, if $j=0$, these modes are degenerate and
must not be counted twice. Hence it is necessary to introduce a degeneracy
factor, $d(j)=1/2\,,j=0$ while $d(j)=1\,,j>0$.

The factoring, S$^3/\Ga$, does not alter the positive--negative split and
the eigenvalues are still as in (\peq{spec}) except that the range of $n$ is
modified. The degeneracy will contain this information.

The deck group, $\Ga$, has left and right actions with typical elements,
$\ga_L$ and $\ga_R$. The degeneracy takes the SO(4) character form,
[\pref{dowded}],
  $$
   d(L,J)={1\over|\Ga|}\sum_{\ga_L,\ga_R}\chi^{(L)}
   (\ga_L)\,\chi^{(J)}(\ga_R)\,,
   \eql{degen2}
  $$
in terms of the SU(2) characters,
  $$
  \chi^{(L)}(\ga)={\sin (2L+1)\th_\ga\over\sin\th_\ga}\,,
  $$
where $\th_\ga$ is the `radial' angular coordinate labelling the group
element, $\ga$.

The freedom allowed by the non--trivial fundamental group to twist the field
by a representation, ${\rm Hom}\,(\Ga,U(1))$, has not been incorporated
here. This would be necessary if the torsion were under consideration.

The spectral data is now combined into a \zf\ on S$^3/\Ga$, \cf
[\pref{DandB}], by adding the positive and negative parts. (Subtraction
would give the $\eta$ invariant).

Trigonometry gives,
  $$
  \ze_3(s)={a^{2s}d(j)\over|\Ga|}
  \sum_{\al,\be}{2\over\cos\be-\cos\al}
  \sumdasht{n=j}{\infty} {1\over n^{2s}}\,\big(\cos n\be\cos j\al-
  \cos n\al\cos j\be\big)\,,
  \eql{zeta1}
  $$
where
  $$
  \al=\th_R+\th_L\,,\quad \be=\th_R-\th_L\,.
  $$

From this point on, the expressions do not apply to spin--half  because the
summation variable has been shifted by $j$ to reach (\peq{zeta1}).

Initially, I will proceed without placing the analysis in a wider context,
\ie just from a pedestrian calculational viewpoint.

As in [\pref{DandB}], the \zf\ is written in terms of the simplest Epstein
\zf, defined by
  $$
  Z\bigg|{g\atop h}\bigg|(s)=\sum_{m=-\infty}^\infty |m+g|^{-s}\,e^{2\pi
  imh}\,,
  \eql{epzet}
  $$
with the understanding that, if $g=0$, the $m=0$ term is omitted. Then,

  $$
 \ze_3(s)={a^{2s}d(j)\over|\Ga|}\sum_{\al,\be}{1\over\cos\be-\cos\al}
 \bigg(\cos j\al\,Z\bigg|{0\atop\be/2\pi}\bigg|(2s)-\cos j\be\,Z\bigg|{0\atop
 \al/2\pi}\bigg|(2s)\bigg)\,.
 \eql{zeta2}
  $$

A remark of possible interest is that, in our previous work,
[\pref{dowsut}], the \zf\ on one--sided (homogeneous) lens spaces was
reduced to a \zf\ on a factored {\it two}--sphere. In the case under study
here, the Epstein function, (\peq{epzet}), is a twisted \zf\ on the {\it
one}--sphere, if $g=0$.

I also note that $\ze_3(0)=0$ for spin--0 but $\ze_3(0)=1$ for spin--1. This
will come up later.

The advantage of the Epstein expression is the existence of a functional
relation that allows (\peq{zeta2}) to be replaced by the image form,
  $$\eqalign{
 \ze_3&(s)={a^{2s}d(j)\over|\Ga|}\pi^{2s-1/2}{\Ga(1/2-s)\over\Ga(s)}\cr
 &\times\sum_{\al,\be}{1\over\cos\be-\cos\al}
 \bigg(\cos j\al \,Z\bigg|{\be/2\pi\atop 0}\bigg|(1-2s)-\cos j\be\, Z\bigg|{
 \al/2\pi\atop 0}\bigg|(1-2s)\bigg)\,.
 }
 \eql{zeta3}
  $$

So far everything has been for a general $\Ga$. The case of lens spaces,
$L(q;\la_1,\la_2)$, is covered by the choice of angles
  $$
  {\al\over2\pi}={ p\nu_1\over q}\,,\quad {\be\over2\pi}={p\nu_2\over q}\,,
  \eql{angles1}
  $$
where $p,\,=0,\ldots,q-1\,,$ labels $\ga$. $\nu_1$ and $\nu_2$ are integers
coprime to $q$, with $\la_1$ and $\la_2$ their mod $q$ inverses.

By an appropriate selection of a $q$-th root of unity, it is possible to set
$\nu_1=1$, \ie\ $\la_1=1$, without loss of generality. Any pair,
$(\nu_1,\nu_2)$, can be reduced to $(1,\nu)$ by multiplying through by the
mod $q$ inverse of $\nu_1$.

The simple, one--sided lens space, $L(q;1,1)$, corresponds to setting
$\nu=1$ so that $\th_L=0$, $\th_R=2\pi p/q$ and the \zf\ becomes a
derivative of an Epstein function. This is not the case I am interested in
just now. Indeed, if the method to be presented later is to run smoothly, it
is necessary that the denominator, $\cos\be-\cos\al$, should never be zero.
This puts conditions on $q$ and $\nu$ which can be simply, but not uniquely,
satisfied by choosing $q$ to be odd prime, when all $\nu$ from $2$ to $q-2$
are covered. I will do this from now on for organisational convenience.

The $p=0$ value corresponds to the identity element of $\Ga$, and is best
separated. For the two spins, the identity \zfs\ are,
  $$\eqalign{
  \ze^{id}_3(s)&=a^{2s}{1\over|\Ga|}\,\ze_R(2s-2)\,,\quad j=0\cr
  \ze^{id}_3(s)&=a^{2s}{2\over|\Ga|}\,\big(\ze_R(2s-2)-\ze_R(2s)
  \big)\,,\quad j=1\,.\cr
  }
  \eql{idzet}
  $$
These differ from the full sphere expressions only by the $1/|\Ga|$ volume
factor.

\section{3. \bf Lens space determinants.}

On differentiating (\peq{zeta2}) at $s=0$ one encounters the right--hand
side bracket evaluated at a point where the $Z$'s have a pole. For spin--0
these cancel, but not for spin--1 when they will combine with the
$1/\Ga(s)$. There is no problem with this, but I proceed in an alternative
fashion eliminating the pole using a limiting procedure.

If $g\ne0$, the $Z$ of (\peq{epzet}) has no pole at $s=1$. So I insert a
non--zero $g$ and let $g\to0$ near the end after the differentiation with
respect to $s$. It is necessary to allow for the extra term at $m=0$
introduced in this way. Hence, instead of (\peq{zeta3}),
  $$\eqalign{
   \ze_3(s)=&{a^{2s}d(j)\over|\Ga|}\pi^{2s-1/2}{\Ga(1/2-s)\over\Ga(s)}
   \sum_{\al,\be}{1\over\cos\be-\cos\al}\cr
 &\times\lim_{g\to0}
 \bigg(\cos j\al \,Z\bigg|{\be/2\pi\atop g}\bigg|(1-2s)-\cos j\be\, Z\bigg|{
 \al/2\pi\atop g}\bigg|(1-2s)\bigg)\cr
 \noalign{\vskip10truept}
 &\hspace{*******}+{d(j)\over|\Ga|}\sum_{\al,\be}
 {\cos j\be-\cos j\al\over\cos\be-\cos\al}\lim_{g\to0}
 \bigg({a\over g}\bigg)^{2s}\,.
 }
 \eql{zeta5}
  $$

Now I can employ a formula of Epstein's, valid for $\nu$ and $q$ integral
and coprime, [\pref{Epstein}],
  $$
  Z\bigg|{\nu/q\atop g}\bigg|(1)=-2\sum_{k=0}^{q-1}e^{-2\pi i(k+g)\nu/q}
  \,\log\sin\big(\pi(g+k)/q\big)\,,
  \eql{epform}
  $$
to give
  $$
  \lim_{g\to0}Z\bigg|{\nu/q\atop g}\bigg|(1)
  =-2\sum_{k=1}^{q-1}e^{-2\pi ik\nu/q}
  \,\log\sin\big(\pi k/q\big)-2\lim_{g \to0}\log\sin(\pi g/q)\,,
  \eql{zelim}
  $$
which is a real quantity and so the exponential can be replaced by a cosine
(set $k\to q-k$).

I note that Epstein's derivation of (\peq{epform}) assumes only the standard
summation,
  $$
  Z\bigg|{0\atop h}\bigg|(1)=-2\log(2\sin\pi h)\,,
  $$
and also that it can be used to streamline some of Ray's algebra,
[\pref{Ray}].

Since the bracket on the right--hand side of (\peq{zeta3}) is now  finite at
$s=0$, in evaluating the derivative at 0, $\ze_3'(0)$, one needs
differentiate only the $1/\Ga(s)$ factor,

  $$\eqalign{
 \ze'_3(0)=-2{d(j)\over q}
 \sum_{p=1}^{q-1}\sum_{k=1}^{q-1}&{\displaystyle
 {\cos{2\pi j\nu p\over q}\cos{2\pi pk\over q}
  -\cos{2\pi j p\over q}\cos{2\pi\nu pk\over q}
 \over\cos{2\pi p\over q}-\cos{2\pi\nu p\over q}}}
\log\sin {\pi k\over q}\cr \noalign{\vskip10truept} +&2{d(j)\over q}
\sum_{p=1}^{q-1}{\displaystyle
 {\cos{2\pi j p\over q} -\cos{2\pi j\nu p\over q}
 \over\cos{2\pi p\over q}-\cos{2\pi\nu p\over q}}}
 \log{\pi a\over q}+\ze^{id\,'}_3(0)\,.
 }
 \eql{zeta4}
  $$
The second term on the right is the residual effect of the `zero mode' for
spin--one and the last term is the contribution of the identity element from
(\peq{idzet}).

The appearance of the radius, $a$, in the spin--1 expression reflects the
non--vanishing of $\ze_3(0)$ and the resulting scaling dependence.

I write (\peq{zeta4}) cosmetically as, after a few cancellations,
  $$\eqalign{
 \ze'_3(0)&=-\sum_{k=1}^{q-1}A_k(0)\log\sin {\pi k\over q}
 +{2\over q}\ze'_R(-2)\,,\quad j=0\cr
 \ze'_3(0)&=-\sum_{k=1}^{q-1}A_k(1)\log\sin {\pi k\over q}
 +{4\over q}\ze'_R(-2)+2\log(2\pi a)
 -{2(q-1)\over q}\log 2q\,,\quad j=1\cr
 }
 \eql{zeta6}
  $$
where the (rational) coefficients, $A_k(j)$, are defined by
  $$
  A_k(j)=2{d(j)\over q}\sum_{p=1}^{q-1}{\displaystyle
 {\cos{2\pi j\nu p\over q}\cos{2\pi pk\over q}
  -\cos{2\pi j p\over q}\cos{2\pi\nu  pk\over q}
 \over\cos{2\pi p\over q}-\cos{2\pi\nu p\over q}}}\,.
\eql{aks}
  $$

I also remark that the same result, (\peq{zeta4}), follows without a special
limit by using the remainder at the pole in the Epstein \zf. This is a
dilogarithm, $\psi$, function, an equivalent statement being,
  $$
  Z'\bigg|{0\atop h}\bigg|(0)=-\ga-\log2\pi-\psi\big({h\over2\pi}\big)-
  \psi\big(1-{h\over2\pi}\big)\,,
  \eql{dif0}
  $$
which could be used directly in (\peq{zeta2}).

One can then employ Gauss' famous formula for $\psi(p/q)$, or better, a
formula that appears during a proof of this relation, [\pref{jensen}] p.146,
which I reproduce, (see also [\pref{AAR}] p.13),
  $$
  \psi\big({p\over q}\big)+\psi\big(1-{p\over q}\big)=-2\ga-2\log2+
  \sum_{k=1}^{q-1}\cos\big({2\pi p k\over q}\big)\log2\sin{\pi q\over k}\,.
  \eql{gff}
  $$
One could also take the attitude that using the Epstein equation,
(\peq{epform}) provides a neat proof of (\peq{gff}), and hence of Gauss'
formula.

The  functional determinants are conventionally defined by
${\caD}_q(j)=e^{-\ze_3'(0)}$.
\begin{ignore}
  $$\eqalign{
  \caD_q(0)&=e^{-2\ze'_R(0)/q}\,\,
\prod_{k=1}^{q-1}\bigg(\sin{\pi k\over q}\bigg)^{A_k(0)} \cr
 \caD_q(1)&={1\over(2\pi a)^2}\,(2q)^{2(q-1)/q}\,e^{-4\ze'_R(0)/q}
\,\, \prod_{k=2}^{q-1}\bigg(\sin{\pi k\over q}\bigg)^{A_k(1)}\,. \cr
  }
  \eql{dets}
  $$
  $$\eqalign{
  D_q(0)&=e^{-2\ze'_R(-2)/q}\,\,
\prod_{k=1}^{[(q-1)/2]}\bigg(\sin^2{\pi k\over q}\bigg)^{A_k(0)} \cr
 D_q(1)&={1\over(2\pi a)^2}\,(2q)^{2(q-1)/q}\,e^{-4\ze'_R(-2)/q}
\,\, \prod_{k=2}^{[(q-1)/2]}\bigg(\sin^2{\pi k\over q}\bigg)^{A_k(1)}\,, \cr
  }
  \eql{dets2}
  $$
\end{ignore}
A convenient quantity is the ratio $R_q$, to the $q$-th root of the full
sphere determinant,
  $$\eqalign{
   R_q(0)={\caD_q(0)\over \big(\caD_1(0)\big)^{1/q}}&=\prod_{k=1}^{[(q-1)/2]}
   \bigg(\sin^2{\pi k\over q}\bigg)^{A_k(0)}\cr
   R_q(1)={\caD_q(1)\over \big(\caD_1(1)\big)^{1/q}}&=
   \bigg({q\over\pi a}\bigg)^{2(1-1/q)}
   \,\,\prod_{k=1}^{[(q-1)/2]}
   \bigg(\sin^2{\pi k\over q}\bigg)^{A_k(1)}\cr
   }
   \eql{rats1}
  $$
where the symmetry $A_{q-k}(j)=A_k(j)$ has been used to fold the product.

In any specific case these quantities can be computed. The first
non--trivial $q$ for which the formula applies is $q=5$, ($\nu=2,3$). This
is because for $q=3$, the possible value of $\nu=2$ is such that $\nu=1$ mod
$q$ and we are back to the one--sided (homogeneous or diagonal) case when a
different method is needed. The method does not work for $q=6$  so the next
example is $q=7$, $(\nu=2,3,4,5)$.

With this in mind, the particular, $\nu$--independent values of the $A_k$,
  $$\eqalign{
  A_1(0)&=1-{1\over q}\,\,,\,\,\,A_2(0)=-{4\over q}\,\,,\,\,\,
  A_3(0)=1-{9\over q}\cr
  A_1(1)&=0\,\,,\,\,\,A_2(1)={2(q-3)\over q}\,.\cr
}
  \eql{avals}
  $$
can be used to rewrite (\peq{rats1}),
  $$\eqalign{
  R_q(0)&={\bigg(\sin^2{\pi \over q}\bigg)^{(q-5)/ q}
 \bigg( \sin^2{3\pi \over q}\bigg)^{(q-9)/ q}
  \over \bigg(4\sec^2{\pi \over q}\bigg)^{{4/ q}}}
\prod_{k=4}^{[(q-1)/2]}\bigg(\sin^2{\pi k\over q}\bigg)^{A_k(0)}\cr
  }
  \eql{rats2}
  $$
for $q>5$ and
  $$\eqalign{
 R_q(1)&=\bigg({q\over\pi a}\bigg)^{2(1-1/q)}
 \bigg(\sin^4{2\pi \over q}\bigg)^{(q-3)/q}
\prod_{k=3}^{[(q-1)/2]}\bigg(\sin^2{\pi k\over q}\bigg)^{A_k(1)} \,.\cr
  }
  \eql{rats3}
  $$
for $q>3$. The algebraic prefactor is the zero mode effect.

The above expressions are useful numerically and I treat them, for now,
purely in this light, as I do equation (\peq{aks}) for the $A_k$'s. The next
section has some further analysis of the $A_k$ quantities and an alternative
route to the equations.

I give a few low values and plot a graph of $W=-\log R_{29}(0)$ against
$\nu$ in the spin--0 case,
  $$\eqalign{
  R_5(0)&=\bigg({47-21\sqrt 5\over2}\bigg)^{1/5}\approx0.46304135
  ,\quad \nu=2,3\cr
  \noalign{\vskip5truept}
  &\approx0.3681520,\hspace{***************} \nu=1,4\cr
  \noalign{\vskip10truept}
  R_7(0)&\approx0.3212271,\hspace{***************}\nu=2,3,4,5\cr
  &\approx0.1679911,\hspace{***************}\nu=1,6\,.
  }
  $$
\newpage

\vglue.8truein
\input epsf
\epsfbox{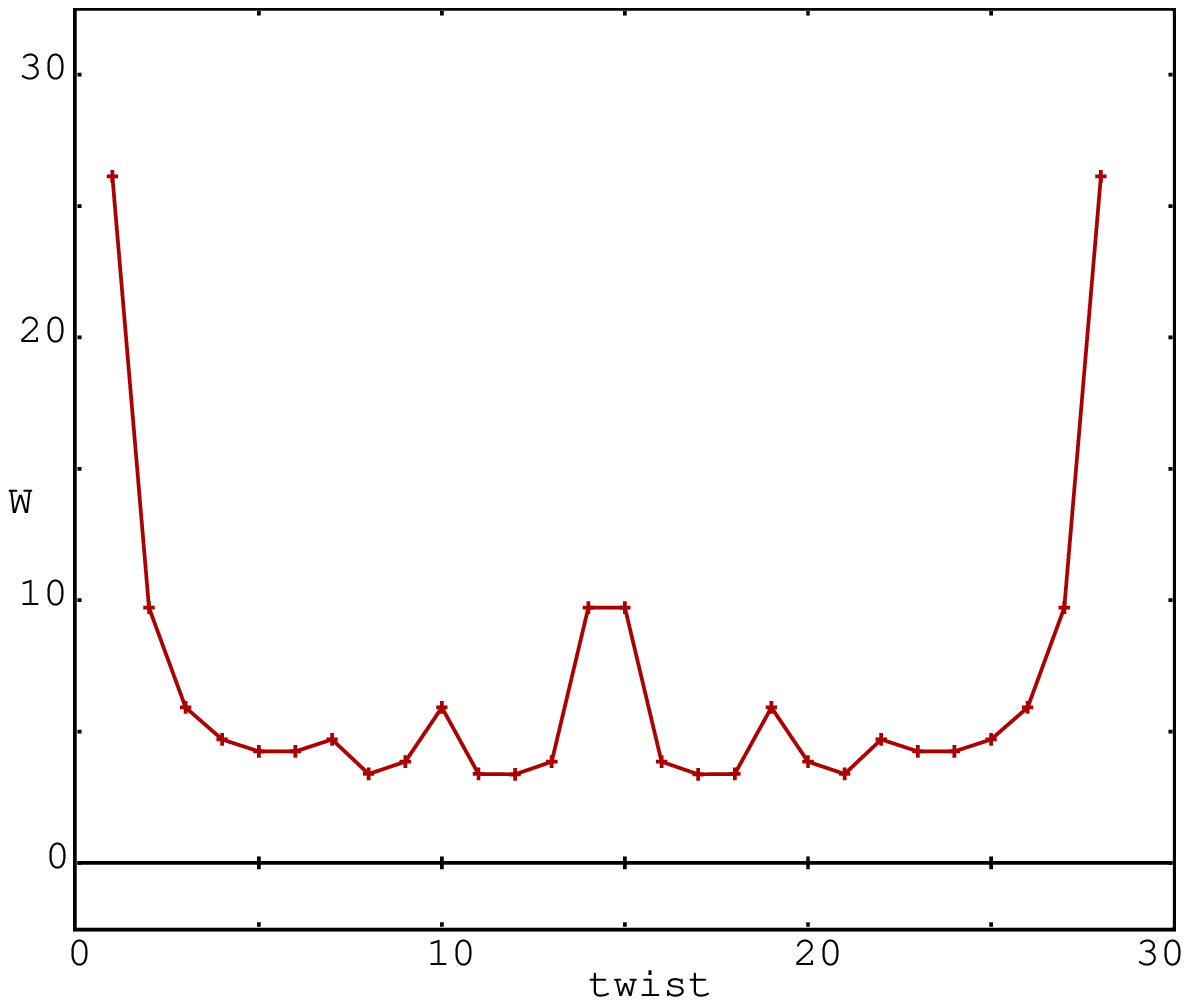} \vskip10truept \begin{narrow} {fig1. $W=-\log R_{29}$ for
conformal scalars on a lens space of order 29

for  twistings $\nu$ or, equivalently, $\la$, from 1 to 28}, ($\la\nu=1$ mod
$29$).\end{narrow}

\vskip30truept

The $\nu=1$ and $\nu=28$ values have been included for completeness. They
were calculated from the expressions developed in section 5.
\section{4. \bf Formal elaboration.}
In tune with my general policy of developing alternative, sometimes
equivalent, approaches, I return to the starting form of the lens space \zf,
(\peq{zeta1}), which, now taking in the definition (\peq{aks}), reads,
  $$
  \ze_3(s)=\ze^{id}_3(s)+a^{2s}\sum_{n=1}^\infty{A_n(j)\over n^{2s}}\,.
\eql{totzet}
  $$
The second term, the non-identity part, is a Dirichlet series.

The $A_n$'s, are obviously periodic from (\peq{aks}),
  $$
  A_{n+q}(j)=A_n(j)\,,
  \eql{ppa}
  $$
and it is traditional in such cases to break up the sum according to
residue--$q$ classes, setting $n=Nq+k$, where $N\in \oZ$ and $0\le k<q$.

From (\peq{aks}),
  $$
  A_0(0)=0,\,\quad A_0(1)=2/q-2\,.
  \eql{0val}
  $$
Since $A_n=A_k$ by (\peq{ppa}), this means that $k$ can be arranged  to run
from $1$ to $q-1$.

The
non--identity \zfs, which are the major technical problem, then become, in
standard fashion,
  $$\eqalign{
  \ze^{nonid}_3(s)&=a^{2s}A_0(j)\sumdasht{N=0}{\infty}{1\over(Nq)^{2s}}+a^{2s}
  \sum_{k=1}^{q-1}A_k(j)\sum_{N=0}^\infty{1\over (Nq+k)^{2s}}\cr
  &=(a/q)^{2s}\bigg[A_0(j)\ze_R(2s)+\sum_{k=1}^{q-1}A_k(j)\ze_R(2s,k/q)
  \bigg]\cr
  &=(a/q)^{2s}\bigg[A_0(j)\ze_R(2s)+\sum_{k=1}^{(q-1)/2}A_k(j)
  \bigg (\ze_R(2s,k/q)+\ze_R(2s,1-k/q)\bigg)\bigg]\cr
  &=(a/q)^{2s}\bigg[A_0(j)\ze_R(2s)+\sum_{k=1}^{(q-1)/2}A_k(j)
  Z\bigg|{k/q\atop0}\bigg|(2s)\bigg]\,,
  }
  \eql{zeta7}
  $$
which constitutes, perhaps, a neater expression than the ones in sections 3
and 4.

The reflection symmetry, from (\peq{aks}),
  $$
  A_{q-k}(j)=A_k(j)\,,
  \eql{sym}
  $$
has been used in (\peq{zeta7}), and earlier.

From general principles, \eg\ [\pref{MandP}], there are no poles in the
non--identity \zf, for fixed point free actions, $\Ga$. From the behaviour
at $s=1/2$ of (\peq{zeta7}) this implies the sum rules,
  $$
  \sum_{k=0}^{q-1}A_k(j)=0\,,
  \eql{tsums}
  $$
which, together with (\peq{0val}) and $\ze_R(0,w)=1/2-w$, yield, correctly,
the values, $\ze_3(0)$, of the {\it total} \zf\ as 0 and 1 for $j=0$ and
$j=1$, respectively.

The local isometry of S$^3$ and S$^3/\Ga$ can be further exploited through
the small--time expansion of the heat--kernel, the coefficients of which are
integrals over local geometrical invariants. They are thus related simply by
a volume $1/|\Ga|$ factor and are determined by the identity \zfs,
(\peq{idzet}). The non--identity contributions must vanish. (Actually, in
the present case, only the first, volume, term in the expansion exists
anyway.) The coefficients are proportional to $\ze_3(-n)$, $n\in\oZ$, and
so, from (\peq{zeta7}), setting $\ze_3^{nonid}(-n)$ to zero, there follows
the sum rules,
  $$
  \sum_{k=1}^{q-1} A_k(j)\,B_{2n+1}(k/q)=0\,.
  \eql{sr}
  $$

In fact, this is only a check because the expressions on the left vanish
identically in view of the reflection symmetry, (\peq{sym}), of the $A_k$
and that of the Bernoulli polynomials,
  $$
  B_n(1-x)=(-1)^n\,B_n(x)\,.
  $$
There is, however, some valuable tactical information that can be extracted
from (\peq{sr}). For example, set $n=0$ and make use of (\peq{tsums}). Then
the sum rules reduce to the moments,
  $$
  \sum_{k=0}^{q-1}k\,A_k(0)=0,\quad \sum_{k=0}^{q-1}k\,A_k(1)=
  {1\over2}\,(q-1)\,.
  \eql{moms}
  $$

The fact that the coefficient of $1/(n/a)^{2s}$ in the \zf\ is a degeneracy,
and hence integral, gives some general information about the $A_n$'s. For
spin--0, the identity part of the \zf\ contributes a (non-periodic) factor
of $n^2/q$ to the (total) degeneracy, hence,
  $$
  A_n(0)=D_n-{n^2\over q}\,,\,(D_n\in\oZ),
  \quad {\rm or}\,\,\,\,qA_n(0)+n^2=0,\,\,{\rm mod}\,\,q\,.
  \eql{propas}
  $$

The values (\peq{avals}) fit this pattern. The degeneracy, $D_n$, depends on
$q$ and the twisting, $\nu$, generally. As a function of $n$, the image
part, $A_n$, oscillates about zero and for large eigenvalues $D_n$ is
dominated by the $n^2$ term, which is the full sphere value divided by $q$,
a volume factor \footnote{The $n^2/q$ is the Weyl term. Using the sum rule,
a sort of discrete eigenvalue counting function is $N_M=\sum_{n=1}^{M}D_n$,
where $M=m(q-1)$, giving $N_M\to M^3/3q$ as $M\to\infty$. $M^2/a^2$ is the
eigenvalue, $\la$, and we see that for large $M$, $N_M\to
\la^{3/2}|\man|/6\pi^2$, the Weyl asymptotic law. Such a crude argument
works only for this leading behaviour}.

As a further check, one can also evaluate the Casimir energy on the lens
space Einstein Universe, T$\times$S$^3/\Ga$,  this way. The values agree, as
they must, with our earlier ones, [\pref{DandB, dowded}]. Non essential
technical comments about this are therefore relegated to Appendix B.

My main objective in this paper is the derivative, $\ze_3'(0)$, and thence
the determinant. This is easy to find by the approach of this section using
Lerch's formula,
  $$
  \ze_R'(0,w)=\log\big(\Ga(w)/\sqrt{2\pi}\big)\,,
  \eql{lerch}
  $$
as in Ray's derivation of the torsion, [\pref{Ray}], or in the evaluation of
the class number for quadratic forms of positive discriminant.

The answers are those given earlier, \eg\ in (\peq{rats1}). In showing this,
one needs again the sums, (\peq{tsums}).

The main point I wish to make now is that, if $q$ is odd prime, the
determinant formulae in (\peq{rats1}) can be written in terms of units,
$\ep_k$, of the $q$-th cyclotomic number field defined by
  $$
  \ep_k=\bigg({(1-\ze^k_q)(1-\ze^{-k}_q)\over(1-\ze_q)
  (1-\ze^{-1}_q)}\bigg)^{1/2}=
  {\sin(\pi k/q)\over\sin(\pi/q)},\quad k=2,\ldots (q-1)/2\,,
  \eql{cu}
  $$
where $\ze_q=\exp(2\pi i/q)$ is a primitive $q$-th root. See Hilbert's
Zahlbericht, [\pref{Hilbert}], and also Borevich and Shafarevich,
[\pref{BandS}], p.360. Franz, [\pref{franz}], uses the square of $\ep$.

Using (\peq{tsums}) and (\peq{avals}), the results are,
  $$\eqalign{
  R_q(0)&=\prod_{k=2}^{(q-1)/2}\ep^{2A_k(0)}_k\cr
  R_q(1)&=\bigg({q\sin\pi/q\over\pi a}\bigg)^{2(1-1/q)}\,
  \prod_{k=2}^{(q-1)/2}\ep^{2A_k(1)}_k\,,
  }
  \eql{unitf}
  $$
where $qA_k(j)\in\oZ$.

The sum rule, (\peq{tsums}), plays a simple but vital role in the derivation
of these formulae.

Because of the differing powers, $A_k$, these expressions are not invariant
under all conjugations of the field.

It is a theorem that the product of units is a unit and also that a rational
power of a unit is a unit, \eg\ [\pref{Hancock2}] \S\S\ 90, 91, 105. Hence,
formally, the ratios in (\peq{unitf}) are cyclotomic units (up to a factor
for spin--one).

I remark that the corresponding spin--1/2 formula is,
  $$
  R_q(1/2)=\prod_{k=1,3,\ldots}^{2q-1}{\ol\ep}^{\,\,A_k(1/2)}_k\,,
  \eql{unitfsp}
  $$
where $\ol\ep_k$ are $2q$--th cyclotomic units, and the $A_k(1/2)$ are the
(partial) degeneracies arising from the non--identity actions. Their form is
given in [\pref{dowded}] and again we have,
  $$
  \sum_{k=1,3,\ldots}^{2q-1}A_k(1/2)=0\,,
  $$
corresponding to $\ze_3(0)=0$ for massless spin--half.
\section{5. \bf Return to the homogeneous case.}

In [\pref{dowsut}], the homogeneous determinants were computed using an
expression for the lens space \zf\ in terms of that on the orbifolded
two--sphere. In this paper, for variety, I wish to take the spin--0,
one--sided Epstein form, [\pref{DandB}],
  $$
  \ze_3(s)=-{a^{2s}\over2|\Ga|}\sum_\ga{1\over\sin\th_\ga}{\pa\over\pa\th_\ga}
  Z\bigg|\matrix{0\cr\th_\ga/2\pi}\bigg|(2s)\,,
  \eql{ep1}
  $$
further to obtain an alternative expression.

There may be a certain amount of
overkill in this, but I have found it helpful to have a number of equivalent
expressions to hand, if only for numerical peace of mind. In view of the
many identities and relations for the \zfs, Bernoulli polynomials \etc, one
should expect several versions of the same thing.

As before, the image form is, [\pref{DandB}],
  $$
  \ze_3(s)=-{a^{2s}\over2|\Ga|}\pi^{2s-1/2}{\Ga(1/2-s)\over\Ga(s)}
\sum_\ga{1\over\sin\th_\ga}{\pa\over\pa\th_\ga}
  Z\bigg|\matrix{\th_\ga/2\pi\cr0}\bigg|(1-2s)\,,
  \eql{ep2}
  $$
which can be re--expressed in terms of the more familiar Hurwitz \zf,
  $$
  \ze_3(s)={\pi^{2s-3/2}\over 2|\Ga|}{\Ga(3/2-s)\over\Ga(s)}\sum_\ga
{1\over\sin\th_\ga}
\bigg(\ze_R\big(2-2s,{\th_\ga\over2\pi}\big)
-\ze_R\big(2-2s,1-{\th_\ga\over2\pi}\big)\bigg)\,.
\eql{zeh}
  $$

This general formula can be used to calculate any quantity required. There
is, however, a slight awkwardness whenever $\th_\ga=\pi$, since both
numerator and denominator in the summand vanish. This happens in even lens
spaces in which case this term, as well as the identity one, is treated
separately. This can be done by extracting the $\oZ_2$ \zf\ scaled by a
volume factor.

Hence, the formula suitable for even $q$ is,
  $$\eqalign{
  \ze_3(s)&=a^{2s}{2\over q}(1-2^{2-2s})\ze_R(2s-2)\cr
  &+{\pi^{2s-3/2}\over q}{\Ga(3/2-s)\over\Ga(s)}\sum_{p=1}^{q/2-1}
{1\over\sin(2\pi p/q)}
\bigg(\ze_R\big(2-2s,{p\over q}\big)
-\ze_R\big(2-2s,1-{p\over q}\big)\bigg)\,,
  }
\eql{zeven}
  $$
Further manipulations would take us to the expressions in
[\pref{dowsut,ChandD}].

Sometimes it is not necessary to make this step. For example, when
evaluating the Casimir energy this way, by setting $s=-1/2$ in (\peq{zeh}),
the Hurwitz functions are evaluated at an odd argument and combine to a sum
that gives a trigonometric result which cancels against the $\sin\th$ on the
bottom. The previous answers in terms of sums of cosecants,
[\pref{DandB,dowsut}], can be obtained in this very roundabout, dual
fashion. I offer some further remarks in Appendix A.

For completeness, the odd $q$ formula is,
  $$\eqalign{
  \ze_3(s&)={a^{2s}\over q}\,\ze_R(2s-2)\cr
  &+{\pi^{2s-3/2}\over q}{\Ga(3/2-s)\over\Ga(s)}\sum_{p=1}^{(q-1)/2}\!\!
{1\over\sin(2\pi p/q)}
\bigg(\ze_R\big(2-2s,{p\over q}\big)
-\ze_R\big(2-2s,1-{p\over q}\big)\bigg)\,.
  }
\eql{zodd}
  $$

The derivatives at $0$, $\ze_3'(0)$, follows more or less as before,
  $$\eqalign{
  \ze'_3(0)=-{a^{2s}\over2\pi^2 q}\,\ze_R(3)
  +{1\over 2\pi q}\sum_{p=1}^{(q-1)/2}
{1\over\sin(2\pi p/q)}\bigg(\ze_R\big(2,{p\over q}\big)
-\ze_R\big(2,1-{p\over q}\big)\bigg)\,,
  }
\eql{dodd}
  $$
for odd $q$ and, for even,
  $$\eqalign{
  \ze'_3(0)=3{a^{2s}\over\pi^2 q}\,\ze_R(3)
  +{1\over 2\pi q}\sum_{p=1}^{q/2-1}
{1\over\sin(2\pi p/q)}\bigg(\ze_R\big(2,{p\over q}\big)
-\ze_R\big(2,1-{p\over q}\big)\bigg)\,,
  }
\eql{deven}
 $$
which constitute decent, numerically amenable forms.

\section{5. \bf Discussion}
The main results of this short paper are the explicit expressions
(\peq{rats1}) and (\peq{unitf}). These yield the determinants on the lens
space, $L(q;1,\nu)$, with a restriction on the coprime $q$ and $\nu$ which
can be met, sufficiently, by restricting $q$ to be odd prime.

Figure 1 graphically summarises the numerics. It obviously exhibits the lens
space homeomorphism $L(q;1,\la)\sim L(q;1,\la')$ when $\la'=\pm\la \,({\rm
mod}\,q)$ and the homeomorphism when $\la'\la=\pm1$ (mod $q$) can also be
observed, \eg, $L(29;1,8)\sim L(29;1,11)$. These equalities explain the
appearance of the graph as a series of extrema and therefore account,
partly, for the startling similarity with the plot of the corresponding
Casimir energy in [\pref{dowded}].

The fact that the image contribution to the determinant is a cyclotomic unit
is presented simply as an amusing formal identity.

\section{\bf Appendix A}

I give some further analysis relevant for the homogeneous lens space
conformal \zf, another method of obtaining which is to apply Plana summation
directly to the sum definition,
  $$
  \ze_3(s,m)=\sum_\Ga \sum_{n=1}^\infty {n\sin n\th_\ga\over\sin\th_\ga
  (n^2+m^2)^s}\,,
  $$
where, because I can, I have added an extra mass--like term. Basic contour
manipulations, \cf [\pref{Dow2}], lead straightforwardly to the `real'
integral for the non--identity, `image', part of the \zf, in the range $s<1$,
  $$
  \ze^{nonid}_3(s,m)=4a^{2s}\sin\pi s\sum_\Ga
 \int_m^\infty {y\,dy\over(y^2-m^2)^s}
  {\sinh y(\th_\ga-\pi)\over\sin\th_\ga\sinh\pi y}\,,\quad 0<\th_\ga<2\pi\,.
  $$

I now set $m=0$ when the integral, [\pref{erdelyi}], [\pref{GandR}],
3.524.1, yields (\peq{zeh}) which can now be extended to all $s$.
Furthermore, [\pref{GandR}] 3.524.10-16, enable one to obtain trigonometric
formulae when $s=-1/2$, \etc\  This is yet another longwinded way of
deriving the cosecant sums in [\pref{DandB,dowsut}].

Useful general formulae to bear in mind are,
  $$
   \int_0^\infty dy\, y^{2r} {\sinh y(\th-\pi)\over\sinh\pi y}
=-{1\over2}{d^{2r}\over d\th^{2r}}\,\cot{\th\over2}
 \eql{inte}
  $$
and
  $$
   \int_0^\infty dy\, y^{2r+1} {\sinh y(\th-\pi)\over\sinh\pi y}
={1\over2}{d^{2r}\over d\th^{2r}}\,\cosec{\th\over2}\,.
  \eql{into}
  $$
Dividing by $\sin\th$, the right--hand side of (\peq{inte}) is a polynomial
in $\cosec^2\th/2$ which can be determined by recursion.\footnote{As a point
of historical interest, the quoted integrals in [\pref{GandR}] are all taken
from the classic compilation by Bierens de Haan, [\pref{Bierens}]. No
references are given in this work, the author referring for  these to vols
IV, V and VIII of the M\'emoires of the Royal Dutch Academy. Only vol IV is
available to me. The lists in these volumes are, apparently, even more
extensive than the mammoth [\pref{Bierens}]!}

Expressed in terms of the series forms, these equations are equivalent to
those of Eisenstein [\pref{Eisenstein}], (see \eg\ Hancock [\pref{Hancock}]
p.32 Ex.5). Defining,
  $$
  (g,x)=\sum_{n=-\infty}^\infty{1\over(x+n)^g}\,,\quad g\in\oZ\,,
  $$
one has the connections,
  $$
  (2g,x)=Z\bigg|{x\atop 0}\bigg|(2g)\,,\quad (2g+1,x)=-{1\over2g}
  {\pa\over\pa x}\,Z\bigg|{x\atop 0}\bigg|(2g)\,.
  $$

The polynomial referred to above is
  $$
  (2g,x)=\pi^{2g}\sum_{k=1}^g(-1)^{k+1}A_{2g,2k}\cosec^{2k}\pi x\,,
  $$
with the recurrence relation
  $$
  A_{2g+2,2k}={1\over2g(2g+1)}\,\big((2k-1)(2k-2)
  A_{2g,2k-2}+4k^2A_{2g,2k}\big)\,,
  $$
and $A_{2g,2g}=1$. Eisenstein remarks that the coefficients are related
simply to Bernoulli numbers.

Likewise one has the expression
  $$
  (2g+1,x)=\pi^{2g+1}\cos\pi x\sum_{k=1}^g(-1)^{k+1}
  A_{2g+1,2k+1}\cosec^{2k+1}\pi x\,.
  $$

If there are further operations to be performed, such as summing over the
angles, then these polynomials are not necessarily the best way of
proceeding. Leaving things as series is sometimes more economical,
[\pref{DandJ,Jadhav}].
\section{\bf Appendix B}
I elaborate on the computation of the Casimir energy using the \zf\ form,
(\peq{totzet}), and make some relational comments on earlier calculations.

The non-identity contribution is,
  $$\eqalign{
  {1\over2}\ze_3^{nonid}&(-1/2)\cr
&={q\over 2a}\bigg[A_0(j)\ze_R(-1)+\sum_{k=1}^{(q-1)/2}A_k(j)
  \bigg (\ze_R\big(-1,{k\over q}\big)
  +\ze_R\big(-1,1-{k\over q}\big)\bigg)\bigg]\cr
  &=-{q\over 4a}\bigg[A_0(j)B_2+\sum_{k=1}^{(q-1)/2}A_k(j)
  \bigg (B_2\big({k\over q}\big)+B_2\big(1-{k\over q}\big)\bigg)\bigg]\cr
  &=-{q\over 4a}\bigg[A_0(j)B_2+2\sum_{k=1}^{(q-1)/2}A_k(j)
  B_2\big({k\over q}\big)\bigg]\cr
  &=-{q\over 2a}\bigg[{A_0(j)\over12}+\sum_{k=1}^{(q-1)/2}A_k(j)\big({k^2\over q^2}
  -{k\over q}+{1\over6}\big)\bigg]\,,
  }
  \eql{casen1}
  $$
in terms of Bernoulli polynomials.

For simplicity, I carry only the spin--0 results further by using the
vanishing moments (\peq{moms}) to give,
  $$
  {1\over2}\ze_3^{nonid}(-1/2)=-{1\over4q a}\sum_{k=0}^{q-1}k^2\,A_k(0)\,.
  \eql{casen2}
  $$

Evaluation of the $A_k$ purely numerically from (\peq{aks}) yields, for
(\peq{casen1}), or (\peq{casen2}), values in complete agreement with those
computed in [\pref{dowded}] which were there expressed in terms of
generalised Dedekind sums. By summing over $n$ first, [\pref{DandB,dowded}],
one comes quickly to the angle sum form,
  $$
 {1\over2}\ze_3^{nonid}(-1/2)=-{1\over 16qa} \sum_{p=1}^{q-1}\cosec^2\pi
 p/q\,\,\cosec^2\pi p\nu/q\,,
  \eql{angf}
  $$
which was determined in [\pref{dowded}] by means of a recursion technique as
an {\it explicit} quartic polynomial in $q$ whose coefficients were found as
numbers, for any given $\nu$.

Arising from the same ingredients, the angle form (\peq{angf}) must be
derivable from the moment form (\peq{casen2}). Using the finite Fourier
transforms,
  $$\eqalign{
  \sum_{k=0}^{q-1}\cos {2\pi k p \nu\over q}&=0,\quad
   \sum_{k=0}^{q-1}k\cos {2\pi k p \nu\over q}=-{q\over2}\cr
    \sum_{k=0}^{q-1}k^2\cos {2\pi k p \nu\over q}&={q\over2}\,
    \cosec^2{\pi p \nu\over q}-{q^2\over2}\,,
    }
    \eql{ffts}
  $$
on (\peq{aks}), with some trigonometry the desired equivalence can be shown.
These transforms also confirm the properties of the $A_k$'s, (\peq{tsums})
and (\peq{moms}).

At the moment I do not have a means of finding the general form for the
partial degeneracies, $A_k$, and so the present way of calculating the
Casimir energy is no better, apart from novelty, than simply performing
the sum,
(\peq{angf}), numerically for given $q$ and $\nu$. Furthermore, in the
present procedure, $q$ has been assumed prime, for convenience.

For $j=0$, the denominator in (\peq{aks}) can be divided into the numerator
at the expense of introducing a product,
  $$
  A_k(0)=2^{k-1}{1\over q}\sum_{p=1}^{q-1}\prod_{r=1}^{k-1}\bigg(\cos{2\pi
  p\over q}-\cos\big({2\pi p \nu\over q}+{2\pi r\over k}\big)\bigg)\,,
  $$
but there seems to be no advantage in this.
\newpage

\section{\bf References.}
\begin{putreferences}
  \ref{DandA}{Dowker,J.S. and Apps, J.S. \cqg{}{}{}.}
  \ref{Weil}{Weil,A., {\it Elliptic functions according to Eisenstein
  and Kronecker}, Springer, Berlin, 1976.}
  \ref{Ling}{Ling,C-H. {\it SIAM J.Math.Anal.} {\bf5} (1974) 551.}
  \ref{Ling2}{Ling,C-H. {\it J.Math.Anal.Appl.}(1988).}
 \ref{BMO}{Brevik,I., Milton,K.A. and Odintsov, S.D. {\it Entropy bounds in
 $R\times S^3$ geometries}. hep-th/0202048.}
 \ref{KandL}{Kutasov,D. and Larsen,F. {\it JHEP} 0101 (2001) 1.}
 \ref{KPS}{Klemm,D., Petkou,A.C. and Siopsis {\it Entropy
 bounds, monoticity properties and scaling in CFT's}. hep-th/0101076.}
 \ref{DandC}{Dowker,J.S. and Critchley,R. \prD{15}{1976}{1484}.}
 \ref{AandD}{Al'taie, M.B. and Dowker, J.S. \prD{18}{1978}{3557}.}
 \ref{Dow1}{Dowker,J.S. \prD{37}{1988}{558}.}
 \ref{Dow3}{Dowker,J.S. \prD{28}{1983}{3013}.}
 \ref{DandK}{Dowker,J.S. and Kennedy,G. \jpa{}{1978}{}.}
 \ref{Dow2}{Dowker,J.S. \cqg{1}{1984}{359}.}
 \ref{DandKi}{Dowker,J.S. and Kirsten, K.{\it Comm. in Anal. and Geom.
 }{\bf7}(1999) 641.}
 \ref{DandKe}{Dowker,J.S. and Kennedy,G.\jpa{11}{1978}{895}.}
 \ref{Gibbons}{Gibbons,G.W. \pl{60A}{1977}{385}.}
 \ref{Cardy}{Cardy,J.L. \np{366}{1991}{403}.}
 \ref{ChandD}{Chang,P. and Dowker,J.S. \np{395}{1993}{407}.}
 \ref{DandC2}{Dowker,J.S. and Critchley,R. \prD{13}{1976}{224}.}
 \ref{Camporesi}{Camporesi,R. \prp{196}{1990}{1}.}
 \ref{BandM}{Brown,L.S. and Maclay,G.J. \pr{184}{1969}{1272}.}
 \ref{CandD}{Candelas,P. and Dowker,J.S. \prD{19}{1979}{2902}.}
 \ref{Unwin1}{Unwin,S.D. Thesis. University of Manchester. 1979.}
 \ref{Unwin2}{Unwin,S.D. \jpa{13}{1980}{313}.}
 \ref{DandB}{Dowker,J.S.and Banach,R. \jpa{11}{1978}{2255}.}
 \ref{Obhukov}{Obhukov,Yu.N. \pl{109B}{1982}{195}.}
 \ref{Kennedy}{Kennedy,G. \prD{23}{1981}{2884}.}
 \ref{CandT}{Copeland,E. and Toms,D.J. \np {255}{1985}{201}.}
 \ref{ELV}{Elizalde,E., Lygren, M. and Vassilevich,
 D.V. \jmp {37}{1996}{3105}.}
 \ref{Malurkar}{Malurkar,S.L. {\it J.Ind.Math.Soc} {\bf16} (1925/26) 130.}
 \ref{Glaisher}{Glaisher,J.W.L. {\it Messenger of Math.} {\bf18}
(1889) 1.} \ref{Anderson}{Anderson,A. \prD{37}{1988}{536}.}
 \ref{CandA}{Cappelli,A. and D'Appollonio,\pl{487B}{2000}{87}.}
 \ref{Wot}{Wotzasek,C. \jpa{23}{1990}{1627}.}
 \ref{RandT}{Ravndal,F. and Tollesen,D. \prD{40}{1989}{4191}.}
 \ref{SandT}{Santos,F.C. and Tort,A.C. \pl{482B}{2000}{323}.}
 \ref{FandO}{Fukushima,K. and Ohta,K. {\it Physica} {\bf A299} (2001) 455.}
 \ref{GandP}{Gibbons,G.W. and Perry,M. \prs{358}{1978}{467}.}
 \ref{Dow4}{Dowker,J.S. {\it Zero modes, entropy bounds and partition
functions.} hep-th\break /0203026.}
  \ref{Rad}{Rademacher,H. {\it Topics in analytic number theory,}
Springer-Verlag,  Berlin,1973.}
  \ref{Halphen}{Halphen,G.-H. {\it Trait\'e des Fonctions Elliptiques}, Vol 1,
Gauthier-Villars, Paris, 1886.}
  \ref{CandW}{Cahn,R.S. and Wolf,J.A. {\it Comm.Mat.Helv.} {\bf 51} (1976) 1.}
  \ref{Berndt}{Berndt,B.C. \rmjm{7}{1977}{147}.}
  \ref{Hurwitz}{Hurwitz,A. \ma{18}{1881}{528}.}
  \ref{Hurwitz2}{Hurwitz,A. {\it Mathematische Werke} Vol.I. Basel,
  Birkhauser, 1932.}
  \ref{Berndt2}{Berndt,B.C. \jram{303/304}{1978}{332}.}
  \ref{RandA}{Rao,M.B. and Ayyar,M.V. \jims{15}{1923/24}{150}.}
  \ref{Hardy}{Hardy,G.H. \jlms{3}{1928}{238}.}
  \ref{TandM}{Tannery,J. and Molk,J. {\it Fonctions Elliptiques},
   Gauthier-Villars, Paris, 1893--1902.}
  \ref{schwarz}{Schwarz,H.-A. {\it Formeln und Lehrs\"atzen zum Gebrauche..},
  Springer 1893.(The first edition was 1885.) The French translation by
Henri Pad\'e is {\it Formules et Propositions pour L'Emploi...},
Gauthier-Villars, Paris, 1894}
  \ref{Hancock}{Hancock,H. {\it Theory of elliptic functions}, Vol I.
   Wiley, New York 1910.}
  \ref{watson}{Watson,G.N. \jlms{3}{1928}{216}.}
  \ref{MandO}{Magnus,W. and Oberhettinger,F. {\it Formeln und S\"atze},
  Springer-Verlag, Berlin 1948.}
  \ref{Klein}{Klein,F. {\it Lectures on the Icosohedron}
  (Methuen, London, 1913).}
  \ref{AandL}{Appell,P. and Lacour,E. {\it Fonctions Elliptiques},
  Gauthier-Villars,
  Paris, 1897.}
  \ref{HandC}{Hurwitz,A. and Courant,C. {\it Allgemeine Funktionentheorie},
  Springer,
  Berlin, 1922.}
  \ref{WandW}{Whittaker,E.T. and Watson,G.N. {\it Modern analysis},
  Cambridge 1927.}
  \ref{SandC}{Selberg,A. and Chowla,S. \jram{227}{1967}{86}. }
  \ref{zucker}{Zucker,I.J. {\it Math.Proc.Camb.Phil.Soc} {\bf 82 }(1977) 111.}
  \ref{glasser}{Glasser,M.L. {\it Maths.of Comp.} {\bf 25} (1971) 533.}
  \ref{GandW}{Glasser, M.L. and Wood,V.E. {\it Maths of Comp.} {\bf 25} (1971)
  535.}
  \ref{greenhill}{Greenhill,A,G. {\it The Applications of Elliptic
  Functions}, MacMillan, London, 1892.}
  \ref{Weierstrass}{Weierstrass,K. {\it J.f.Mathematik (Crelle)}
{\bf 52} (1856) 346.}
  \ref{Weierstrass2}{Weierstrass,K. {\it Mathematische Werke} Vol.I,p.1,
  Mayer u. M\"uller, Berlin, 1894.}
  \ref{Fricke}{Fricke,R. {\it Die Elliptische Funktionen und Ihre Anwendungen},
    Teubner, Leipzig. 1915, 1922.}
  \ref{Konig}{K\"onigsberger,L. {\it Vorlesungen \"uber die Theorie der
 Elliptischen Funktionen},  \break Teubner, Leipzig, 1874.}
  \ref{Milne}{Milne,S.C. {\it The Ramanujan Journal} {\bf 6} (2002) 7-149.}
  \ref{Schlomilch}{Schl\"omilch,O. {\it Ber. Verh. K. Sachs. Gesell. Wiss.
  Leipzig}  {\bf 29} (1877) 101-105; {\it Compendium der h\"oheren Analysis},
  Bd.II, 3rd Edn, Vieweg, Brunswick, 1878.}
  \ref{BandB}{Briot,C. and Bouquet,C. {\it Th\`eorie des Fonctions
  Elliptiques}, Gauthier-Villars, Paris, 1875.}
  \ref{Dumont}{Dumont,D. \aim {41}{1981}{1}.}
  \ref{Andre}{Andr\'e,D. {\it Ann.\'Ecole Normale Superior} {\bf 6} (1877) 265;
  {\it J.Math.Pures et Appl.} {\bf 5} (1878) 31.}
  \ref{Raman}{Ramanujan,S. {\it Trans.Camb.Phil.Soc.} {\bf 22} (1916) 159;
 {\it Collected Papers}, Cambridge, 1927}
  \ref{Weber}{Weber,H.M. {\it Lehrbuch der Algebra} Bd.III, Vieweg,
  Brunswick 190  3.}
  \ref{Weber2}{Weber,H.M. {\it Elliptische Funktionen und algebraische Zahlen},
  Vieweg, Brunswick 1891.}
  \ref{ZandR}{Zucker,I.J. and Robertson,M.M.
  {\it Math.Proc.Camb.Phil.Soc} {\bf 95 }(1984) 5.}
  \ref{JandZ1}{Joyce,G.S. and Zucker,I.J.
  {\it Math.Proc.Camb.Phil.Soc} {\bf 109 }(1991) 257.}
  \ref{JandZ2}{Zucker,I.J. and Joyce.G.S.
  {\it Math.Proc.Camb.Phil.Soc} {\bf 131 }(2001) 309.}
  \ref{zucker2}{Zucker,I.J. {\it SIAM J.Math.Anal.} {\bf 10} (1979) 192,}
  \ref{BandZ}{Borwein,J.M. and Zucker,I.J. {\it IMA J.Math.Anal.} {\bf 12}
  (1992) 519.}
  \ref{Cox}{Cox,D.A. {\it Primes of the form $x^2+n\,y^2$}, Wiley, New York,
  1989.}
  \ref{BandCh}{Berndt,B.C. and Chan,H.H. {\it Mathematika} {\bf42} (1995) 278.}
  \ref{EandT}{Elizalde,R. and Tort.hep-th/}
  \ref{KandS}{Kiyek,K. and Schmidt,H. {\it Arch.Math.} {\bf 18} (1967) 438.}
  \ref{Oshima}{Oshima,K. \prD{46}{1992}{4765}.}
  \ref{greenhill2}{Greenhill,A.G. \plms{19} {1888} {301}.}
  \ref{Russell}{Russell,R. \plms{19} {1888} {91}.}
  \ref{BandB}{Borwein,J.M. and Borwein,P.B. {\it Pi and the AGM}, Wiley,
  New York, 1998.}
  \ref{Resnikoff}{Resnikoff,H.L. \tams{124}{1966}{334}.}
  \ref{vandp}{Van der Pol, B. {\it Indag.Math.} {\bf18} (1951) 261,272.}
  \ref{Rankin}{Rankin,R.A. {\it Modular forms} CUP}
  \ref{Rankin2}{Rankin,R.A. {\it Proc. Roy.Soc. Edin.} {\bf76 A} (1976) 107.}
  \ref{Skoruppa}{Skoruppa,N-P. {\it J.of Number Th.} {\bf43} (1993) 68 .}
  \ref{Down}{Dowker.J.S. \np {104}{2002}{153}.}
  \ref{Eichler}{Eichler,M. \mz {67}{1957}{267}.}
  \ref{Zagier}{Zagier,D. \invm{104}{1991}{449}.}
  \ref{Lang}{Lang,S. {\it Modular Forms}, Springer, Berlin, 1976.}
  \ref{Kosh}{Koshliakov,N.S. {\it Mess.of Math.} {\bf 58} (1928) 1.}
  \ref{BandH}{Bodendiek, R. and Halbritter,U. \amsh{38}{1972}{147}.}
  \ref{Smart}{Smart,L.R., \pgma{14}{1973}{1}.}
  \ref{Grosswald}{Grosswald,E. {\it Acta. Arith.} {\bf 21} (1972) 25.}
  \ref{Kata}{Katayama,K. {\it Acta Arith.} {\bf 22} (1973) 149.}
  \ref{Ogg}{Ogg,A. {\it Modular forms and Dirichlet series} (Benjamin,
  New York,
   1969).}
  \ref{Bol}{Bol,G. \amsh{16}{1949}{1}.}
  \ref{Epstein}{Epstein,P. \ma{56}{1903}{615}.}
  \ref{Petersson}{Petersson.}
  \ref{Serre}{Serre,J-P. {\it A Course in Arithmetic}, Springer,
  New York, 1973.}
  \ref{Schoenberg}{Schoenberg,B., {\it Elliptic Modular Functions},
  Springer, Berlin, 1974.}
  \ref{Apostol}{Apostol,T.M. \dmj {17}{1950}{147}.}
  \ref{Ogg2}{Ogg,A. {\it Lecture Notes in Math.} {\bf 320} (1973) 1.}
  \ref{Knopp}{Knopp,M.I. \dmj {45}{1978}{47}.}
  \ref{Knopp2}{Knopp,M.I. \invm {}{1994}{361}.}
  \ref{LandZ}{Lewis,J. and Zagier,D. \aom{153}{2001}{191}.}
  \ref{DandK1}{Dowker,J.S. and Kirsten,K. {\it Elliptic functions and
  temperature inversion symmetry on spheres} hep-th/.}
  \ref{HandK}{Husseini and Knopp.}
  \ref{Kober}{Kober,H. \mz{39}{1934-5}{609}.}
  \ref{HandL}{Hardy,G.H. and Littlewood, \am{41}{1917}{119}.}
  \ref{Watson}{Watson,G.N. \qjm{2}{1931}{300}.}
  \ref{SandC2}{Chowla,S. and Selberg,A. {\it Proc.Nat.Acad.} {\bf 35}
  (1949) 371.}
  \ref{Landau}{Landau, E. {\it Lehre von der Verteilung der Primzahlen},
  (Teubner, Leipzig, 1909).}
  \ref{Berndt4}{Berndt,B.C. \tams {146}{1969}{323}.}
  \ref{Berndt3}{Berndt,B.C. \tams {}{}{}.}
  \ref{Bochner}{Bochner,S. \aom{53}{1951}{332}.}
  \ref{Weil2}{Weil,A.\ma{168}{1967}{}.}
  \ref{CandN}{Chandrasekharan,K. and Narasimhan,R. \aom{74}{1961}{1}.}
  \ref{Rankin3}{Rankin,R.A. {} {} ().}
  \ref{Berndt6}{Berndt,B.C. {\it Trans.Edin.Math.Soc}.}
  \ref{Elizalde}{Elizalde,E. {\it Ten Physical Applications of Spectral
  Zeta Function Theory}, \break (Springer, Berlin, 1995).}
  \ref{Allen}{Allen,B., Folacci,A. and Gibbons,G.W. \pl{189}{1987}{304}.}
  \ref{Krazer}{Krazer}
  \ref{Elizalde3}{Elizalde,E. {\it J.Comp.and Appl. Math.} {\bf 118}
  (2000) 125.}
  \ref{Elizalde2}{Elizalde,E., Odintsov.S.D, Romeo, A. and Bytsenko,
  A.A and
  Zerbini,S.
  {\it Zeta function regularisation}, (World Scientific, Singapore,
  1994).}
  \ref{Eisenstein}{Eisenstein,F. {\it J.Math. (Crelle)} {\bf 35} (1847) 153.}
  \ref{Hecke}{Hecke,E. \ma{112}{1936}{664}.}
  \ref{Terras}{Terras,A. {\it Harmonic analysis on Symmetric Spaces} (Springer,
  New York, 1985).}
  \ref{BandG}{Bateman,P.T. and Grosswald,E. {\it Acta Arith.} {\bf 9}
  (1964) 365.}
  \ref{Deuring}{Deuring,M. \aom{38}{1937}{585}.}
  \ref{Guinand}{Guinand.}
  \ref{Guinand2}{Guinand.}
  \ref{Minak}{Minakshisundaram.}
  \ref{Mordell}{Mordell,J. \prs{}{}{}.}
  \ref{GandZ}{Glasser,M.L. and Zucker, {}.}
  \ref{Landau2}{Landau,E. \jram{}{1903}{64}.}
  \ref{Kirsten1}{Kirsten,K. \jmp{35}{1994}{459}.}
  \ref{Sommer}{Sommer,J. {\it Vorlesungen \"uber Zahlentheorie}
  (1907,Teubner,Leipzig).
  French edition 1913 .}
  \ref{Reid}{Reid,L.W. {\it Theory of Algebraic Numbers},
  (1910,MacMillan,New York).}
  \ref{Milnor}{Milnor, J. {\it Is the Universe simply--connected?},
  IAS, Princeton, 1978.}
  \ref{Milnor2}{Milnor, J. \ajm{79}{1957}{623}.}
  \ref{Opechowski}{Opechowski,W. {\it Physica} {\bf 7} (1940) 552.}
  \ref{Bethe}{Bethe, H.A. \zfp{3}{1929}{133}.}
  \ref{LandL}{Landau, L.D. and Lishitz, E.M. {\it Quantum
  Mechanics} (Pergamon Press, London, 1958).}
  \ref{GPR}{Gibbons, G.W., Pope, C. and R\"omer, H., \np{157}{1979}{377}.}
  \ref{Jadhav}{Jadhav,S.P. PhD Thesis, University of Manchester 1990.}
  \ref{DandJ}{Dowker,J.S. and Jadhav, S. \prD{39}{1989}{1196}.}
  \ref{CandM}{Coxeter, H.S.M. and Moser, W.O.J. {\it Generators and
  relations of finite groups} Springer. Berlin. 1957.}
  \ref{Coxeter2}{Coxeter, H.S.M. {\it Regular Complex Polytopes},
   (Cambridge University Press,
  Cambridge, 1975).}
  \ref{Coxeter}{Coxeter, H.S.M. {\it Regular Polytopes}.}
  \ref{Stiefel}{Stiefel, E., J.Research NBS {\bf 48} (1952) 424.}
  \ref{BandS}{Brink and Satchler {\it Angular momentum theory}.}
  \ref{Rose}{Rose}
  \ref{Schwinger}{Schwinger,J.}
  \ref{Bromwich}{Bromwich, T.J.I'A. {\it Infinite Series},
  (Macmillan, 1947).}
  \ref{Ray}{Ray,D.B. \aim{4}{1970}{109}.}
  \ref{Ikeda}{Ikeda,A. {\it Kodai Math.J.} {\bf 18} (1995) 57.}
  \ref{Kennedy}{Kennedy,G. \prD{23}{1981}{2884}.}
  \ref{Ellis}{Ellis,G.F.R. {\it General Relativity} {\bf2} (1971) 7.}
  \ref{Dow8}{Dowker,J.S. \cqg{20}{2003}{L105}.}
  \ref{IandY}{Ikeda, A and Yamamoto, Y. \ojm {16}{1979}{447}.}
  \ref{BandI}{Bander,M. and Itzykson,C. \rmp{18}{1966}{2}.}
  \ref{Schulman}{Schulman, L.S. \pr{176}{1968}{1558}.}
  \ref{Bar1}{B\"ar,C. {\it Arch.d.Math.}{\bf 59} (1992) 65.}
  \ref{Bar2}{B\"ar,C. {\it Geom. and Func. Anal.} {\bf 6} (1996) 899.}
  \ref{Vilenkin}{Vilenkin, N.J. {\it Special functions},
  (Am.Math.Soc., Providence, 1968).}
  \ref{Talman}{Talman, J.D. {\it Special functions} (Benjamin,N.Y.,1968).}
  \ref{Miller}{Miller,W. {\it Symmetry groups and their applications}
  (Wiley, N.Y., 1972).}
  \ref{Dow3}{Dowker,J.S. \cmp{162}{1994}{633}.}
  \ref{Cheeger}{Cheeger, J. \jdg {18}{1983}{575}.}
  \ref{Dow6}{Dowker,J.S. \jmp{30}{1989}{770}.}
  \ref{Dow9}{Dowker,J.S. \jmp{42}{2001}{1501}.}
  \ref{Dow7}{Dowker,J.S. \jpa{25}{1992}{2641}.}
  \ref{Warner}{Warner.N.P. \prs{383}{1982}{379}.}
  \ref{Wolf}{Wolf, J.A. {\it Spaces of constant curvature},
  (McGraw--Hill,N.Y., 1967).}
  \ref{Meyer}{Meyer,B. \cjm{6}{1954}{135}.}
  \ref{BandB}{B\'erard,P. and Besson,G. {\it Ann. Inst. Four.} {\bf 30}
  (1980) 237.}
  \ref{PandM}{Polya,G. and Meyer,B. \cras{228}{1948}{28}.}
  \ref{Springer}{Springer, T.A. Lecture Notes in Math. vol 585 (Springer,
  Berlin,1977).}
  \ref{SeandT}{Threlfall, H. and Seifert, W. \ma{104}{1930}{1}.}
  \ref{Hopf}{Hopf,H. \ma{95}{1925}{313}. }
  \ref{Dow}{Dowker,J.S. \jpa{5}{1972}{936}.}
  \ref{LLL}{Lehoucq,R., Lachi\'eze-Rey,M. and Luminet, J.--P. {\it
  Astron.Astrophys.} {\bf 313} (1996) 339.}
  \ref{LaandL}{Lachi\'eze-Rey,M. and Luminet, J.--P.
  \prp{254}{1995}{135}.}
  \ref{Schwarzschild}{Schwarzschild, K., {\it Vierteljahrschrift der
  Ast.Ges.} {\bf 35} (1900) 337.}
  \ref{Starkman}{Starkman,G.D. \cqg{15}{1998}{2529}.}
  \ref{LWUGL}{Lehoucq,R., Weeks,J.R., Uzan,J.P., Gausman, E. and
  Luminet, J.--P. \cqg{19}{2002}{4683}.}
  \ref{Dow10}{Dowker,J.S. \prD{28}{1983}{3013}.}
  \ref{BandD}{Banach, R. and Dowker, J.S. \jpa{12}{1979}{2527}.}
  \ref{Jadhav2}{Jadhav,S. \prD{43}{1991}{2656}.}
  \ref{Gilkey}{Gilkey,P.B. {\it Invariance theory,the heat equation and
  the Atiyah--Singer Index theorem} (CRC Press, Boca Raton, 1994).}
  \ref{BandY}{Berndt,B.C. and Yeap,B.P. {\it Adv. Appl. Math.}
  {\bf29} (2002) 358.}
  \ref{HandR}{Hanson,A.J. and R\"omer,H. \pl{80B}{1978}{58}.}
  \ref{Hill}{Hill,M.J.M. {\it Trans.Camb.Phil.Soc.} {\bf 13} (1883) 36.}
  \ref{Cayley}{Cayley,A. {\it Quart.Math.J.} {\bf 7} (1866) 304.}
  \ref{Seade}{Seade,J.A. {\it Anal.Inst.Mat.Univ.Nac.Aut\'on
  M\'exico} {\bf 21} (1981) 129.}
  \ref{CM}{Cisneros--Molina,J.L. {\it Geom.Dedicata} {\bf84} (2001)
  \ref{Goette1}{Goette,S. \jram {526} {2000} 181.}
  207.}
  \ref{NandO}{Nash,C. and O'Connor,D--J, \jmp {36}{1995}{1462}.}
  \ref{Dows}{Dowker,J.S. \aop{71}{1972}{577}; Dowker,J.S. and Pettengill,D.F.
  \jpa{7}{1974}{1527}; J.S.Dowker in {\it Quantum Gravity}, edited by
  S. C. Christensen (Hilger,Bristol,1984)}
  \ref{Jadhav2}{Jadhav,S.P. \prD{43}{1991}{2656}.}
  \ref{Dow11}{Dowker,J.S. {\it Spherical Universe topology and the Casimir
  effect} hep-th/0404093.}
  \ref{Zagier}{Zagier,D. \ma{202}{1973}{149}}
  \ref{RandG}{Rademacher, H. and Grosswald,E. {\it Dedekind Sums},
  (Carus, MAA, 1972).}
  \ref{Berndt7}{Berndt,B, \aim{23}{1977}{285}.}
  \ref{HKMM}{Harvey,J.A., Kutasov,D., Martinec,E.J. and Moore,G. {\it
  Localised Tachyons and RG Flows}, hep-th/0111154.}
  \ref{Beck}{Beck,M., {\it Dedekind Cotangent Sums}, {\it Acta Arithmetica}
  {\bf 109} (2003) 109-139 ; math.NT/0112077.}
  \ref{McInnes}{McInnes,B. {\it APS instability and the topology of the brane
  world}, hep-th/0401035.}
  \ref{DandK5}{Dowker,J.S. and Kirsten, Klaus {\it The Barnes zeta function,
   sphere determinants and the Glaisher--Kinkelin--Bendersky constants},
    hep-th/0301143. }
  \ref{dowsut}{Dowker,J.S. {\it Spherical Universe topology and the Casimir
  effect} hep-th/0404093.}
  \ref{dowded}{Dowker,J.S. {\it Spherical Casimir energies and Dedekind
  sums} hep-th/0406113.}
  \ref{GandR}{Gradshteyn,I.S. and Ryzhik,I.M., {\it Tables of Integrals,
Series and Products}, (Academic Press, New York, 1965).}
  \ref{Bierens}{Bierens de Haan,D. {\it Nouvelles tables d'int\'egrales
  d\'efinies}, (P.Engels, Leiden, 1867).}
  \ref{erdelyi}{Erdelyi,A. {\it Tables of integral transforms} Vols. I and
II (McGraw--Hill, New York, 1954).}
  \ref{AAR}{Andrews, G.E., Askey,R. and Roy,R. {\it Special functions}
  (CUP, Cambridge, 1999).}
  \ref{jensen}{Jensen,J.L.W.V. \aom {17}{1915-1916}{124}.}
  \ref{franz}{Franz,W. \jram {173}{1935}{245}.}
  \ref{Hancock2}{Hancock,H. {\it Foundations of the theory of algebraic
  numbers}, Vol.I (Macmillan, New York, 1931).}
 \ref{Hilbert}{Hilbert,D. {\it The theory of algebraic number fields},
translated by I.T.Adamson, (Springer, Berlin, 1998).}
  \ref{BandS}{Borevich,Z.I. and Shafarevich,I.R. {\it Number theory} (Academic
  Press, New York, 1966).}
  \ref{MandP}{Minakshisudaram,S. and Pleijel,A. \cjm{1}{1949}{242}.}
\end{putreferences}

 \bye